# Tuned Vacancy Diffusion by Mn via Anomalous Friedel Oscillations in NiCoFeCrMn High Entropy Alloys

Shaosong Huang[a*], Huaqing Guan[a], Fuyang Tian[b], Chenyang Lu[c*], Qiu Xu[d], Jijun Zhao[a]

[a] Key Laboratory of Materials Modification by Laser, Ion and Electron Beams (Ministry of Education), Dalian University of Technology, Dalian, 116024, China

[b] Institute for Applied Physics, University of Science and Technology Beijing, Beijing, 10083, China

[c] Department of Nuclear Science and Technology, Xi'an Jiaotong University, Xi'an, 710049, China

[d] Institute for Integrated Radiation and Nuclear Science, Kyoto University, Osaka, 5900494, Japan

# Abstract

Vacancy is known to play a critical role in solid state diffusion and determine many properties of the materials. We report a proof-of-principle result on Mn in the role of tuning vacancy diffusion by an anomalous Friedel Oscillations effect in an equivalent NiCoFeCrMn high entropy alloy. The vacancy induced Friedel Oscillations is destroyed by the presence of the first nearest neighboring Mn, opening a channel for long-range influence on the surrounding atoms from the vacancy, resulting in a lower vacancy diffusion barrier energy, despite of the atom type. The effects lead to a remarkable diffusion overlaps of the two type defects, vacancy and interstitial, enhancing their recombination, which can explain well for the experimental results with suppression of vacancy clusters under irradiation. This finding presents new understanding on the atomic diffusion mechanism and thus suggests novel concepts for tuning defect properties in the design of advanced materials with outstanding properties, in a broad field.

The properties of materials are essentially determined by the defects, which govern the kinetics of microstructural changes and processes of mass transport at elevated temperatures. How to link defect dynamics with intrinsic properties is an old question in materials science. Especially, in the field of materials used under irradiation, where a continuous large number of point defects can be created by high energy particle bombardment. The survivability and aggregation of which can form extended defects and degrade the macro-properties of the materials, resulting in the so-called irradiation damage [1]. The degradation is essentially stems from the property difference of vacancy and interstitial [2]. Therefore, to control/minimize the differentiated performance of the point defects and their afterward associates has become the ultimate way for developing advanced materials with high irradiation tolerance. Such as introducing high-densities of interfaces: (1) oxide dispersion-strengthened steels [3,4], (2) nano-grained polycrystalline alloys [5] and so on, to provide more annihilation sites for different type defects with mitigating the irradiation effects as a consequence. In recent years, unlike these strategies, studies have shown that the concentrated solid-solution alloys (CSA) or the high-entropy alloys (HEA) [6] have excellent anti-radiation properties in various temperature regimes through modifying their compositional complexity [7-10].

Studies suggested that HEAs possess "special" irradiation damage resistance due to sluggish diffusion in the form of vacancy clusters [11], dislocation [12], and interstitial clusters [8] with three-dimensional diffusion mode in CSAs versus one-dimensional mode in pure Ni. Recently, reduced mean free path of electrons and poor thermal conductivities on the displacement cascade [13]were also reported for the Ni-based CSA. However, research showed that vacancy diffusion in the CSAs is not necessarily sluggish [14]. The number of alloying elements may not be critical [15], but the specified elements are quite important to the irradiation response [10]. Furthermore, analysis concluded that the 1D to 3D motion of small interstitial clusters do not support the swelling resistance of CSAs but the notion that compositional changes can have a large influence on the formation of extended defects [10]. However, and most importantly, the scientific basis on modifying the intrinsic properties by the elements is still far from understood. Improved understanding of how element-specific chemical complexity affects defect physics is indispensable to interpret the experimental findings and further optimize alloy development among the enormous combinations of high entropy alloys.

Experimental research on point defect dynamics with element discrimination is extremely difficult due to facility limitations. Atomistic investigation of HEAs are critical but complicated by a large number of constituents that entail a combinatorically high computational and human effort [16]. It is also challenging because of the difficulty to sample the chemical disorder. To address these challenges, in this study, a similar atomic environment (SAE) method-based density functional theory (DFT) [17,18] and a large number of statistical atomic sites are performed to avoid the limitations of unique site-to-site atomic environments. A striking phenomenon and its underlying mechanism on Mn in the role of increasing the effective vacancy mobility and enhance the vacancy-interstitial recombination in an equivalent NiCoFeCrMn high entropy alloy will be shown.

The known Cantor-Wu alloy NiCoFeCrMn and NiCoFeCr were selected to investigate the experimental irradiation response. The alloys were prepared as that in the paper [19], and irradiated with 3 MeV $Ni^{2+}$ ions to a fluence of 5 x $10^{16}$ ions/$cm^2$ at 853 K in the Ion Beam Materials Laboratory (IBML) at the University of Tennessee. The relative high irradiation temperature was chosen so that both the onset migration of interstitial and vacancy can occur to check the void swelling response [2]. Ion irradiation doses with a peak value of ~60 displacements per atom (dpa) and stopping range in the samples were computed by SRIM 2013 assuming a displacement threshold energy of 40 eV in Kinchin-Pease option [20]. The damage profile is shown in Fig.1(a). The cross-section bright field TEM images show clearly that after irradiation, the size and number densities of vacancy clusters in NiCoFeCrMn are much smaller and lower than those in NiCoFeCr. The addition of Mn significantly suppresses irradiation-induced void formation to a factor of 12.

To explore the underlying mechanisms, DFT calculations are conducted using the VASP [21-23]. Detailed methodology is described in the supplementary materials [24]. NEB calculations of vacancy and interstitial migration barriers based on different element types in NiCoFeCr and NiCoFeCrMn supercells are performed. The calculated diffusion paths are shown in Fig.S1 to S4. The migration barrier ranges of vacancy and interstitial are summarized in Fig.1(b). Interestingly, rather than NiCoFeCr, a remarkable overlap region in NiCoFeCrMn can be seen. The smaller separations of diffusion between the two type point defects will facilitate their short-range recombination during the cascade process with less defect accumulation, slow the formation of vacancy super-saturations, which is a pre-requisite for void swelling [25]. To correlate the results with experimental, it is necessary to ensure that the energies are available at temperature T. Based on the harmonic transition state theory, when a set of atoms is in equilibrium at temperature T, the average energy available to each degree of freedom in the system is $K_BT/2$, where $K_B$ is the Boltzmann constant [26]. At the irradiation temperature 853K, $K_BT/2$ is roughly 0.03 eV. The average energy barrier for an atom moving from a hollow to a bridge site is 0.3 eV, more than 10 times the amount of average thermal energy. Therefore, the temperature will rarely affect the value of the diffusion barriers. So, the overlap in NiCoFeCrMn alloy is independent of the temperature to some extent. Which, in other words, can explain well for the experimental results with suppression of vacancy clusters in NiCoFeCrMn alloy.

As shown in Fig.1(b), the energy overlap in NiCoFeCrMn alloy is mainly due to the decrease of vacancy diffusion barrier. What makes vacancy easy to diffuse in the alloy? Intuitively, the addition of Mn may be a simple criterion, as it contributes to the enhanced elemental multiformity. As is known, Mn has an atypical and complex electronic structure with a half-filled 3*d*-shell and can exhibit two states with ferromagnetic (FM) and antiferromagnetic (AFM). The insight behind the Mn effect on the vacancy diffusion could be correlated to the magnetic moment and electronic structure properties. In solid, when a vacancy is introduced, the surrounding electrons at the Fermi energy will have a shielding effect on the defect, which is caused by Friedel oscillations (FO) of the electron density distribution due to microscopic electronic volatility [27,28]. FO are spatial modulations of electron spin or charge density that result from a defect or boundary, which was proposed in [29] and experimentally proved recently [30]. The vacancy formation will lead to a local electronic redistribution and induce a depletion of the surrounding element charge, which is Friedel-like and can only involve 1-2nn atoms, as typically shown in Fig.2(a) and (c) for NiCoFeCr. However, in sharp contrast, with the same isosurface of 0.04 e/Å$^3$, the large electron density oscillations can spread as far as more than 6nn in NiCoFeCrMn alloy (Fig.2(b) and (d)). As shown in Fig.3, the vacancy induced FO or electron screen effect is destroyed by the presence of the 1nn Mn. Interestingly, unlike the other four elements, the local electron charge of these Mn atoms, however, do not change much, forming an "opening channel" in between the atoms. The "opening channel" should be attributed to the unique half-filled 3*d* electrons of Mn. The FO is essentially related to the indirect coupling between magnetic moments via the conduction electrons in a metal with the famous Ruderman–Kittel–Kasuya–Yosida (RKKY) interaction potential [31]. Recently, ab-initio calculations [32] on residual resistivity of Cantor-Wu alloys showed that for Mn containing alloys spin channels experience strong disorder scattering due to an electron filling effect. As demonstrated in Fig.S5 the projected density of states (PDOS) of Mn in five 3*d*-orbitals, for all the three 1nn Mn, most of the occupation numbers of the spin are either majority or minority, indicating a large number of empty orbits, which serves as a transmission channel for the influence of vacancy. For Mn2, we can see clearly that the formation of vacancy makes majority-spin part more bigly. Since the local electron transfer between Mn and its surrounding atoms is very small after the generation of vacancy, the electron redistribution of the Mn could be a result of the lowered symmetry due to the introduction of the vacancy [25]. With FO, around the vacancy there is a charge localization cage, The relaxation of the nearest neighbor shells oscillates as compression, tension, compression [33]. Owing to the repulsive electron–electron interaction, it would be difficult for the surrounding atoms to exchange with the vacancy, resulting in a high diffusion barrier. On the other hand, the destruction of the FO by Mn can restore and spread the original lattice distortion effect of the vacancy. The comparison of the diffusion barrier data of the surrounding atoms in the two alloys shown in Fig.2(e) and (f) almost perfectly approve this conclusion, with most of the value are smaller in NiCoFeCrMn alloy than those in NiCoFeCr alloy for the same element type. Furthermore, the local 1nn atomic environment for the diffusing elements would be particularly important, considering the Mn role as "opening channel". TableS1 shows the diffusion routes with 1nn atomic environment of the diffusing elements for two single-vacancy V1 and V2 in NiCoFeCrMn alloy. More than expected, for each vacancy, the diffusion barrier of element is inversely proportional to the number of its 1nn Mn atom to an almost perfect extent, and is more sensitive to the Mn in the diffusion path, consistent with the Mn role as "opening channel".

Note that, the diffusion barrier $E_{mig}$ is indeed the energy difference between the solute atom at the initial and saddle point. The atomic state at the saddle point is truly also important. We choose two different paths for the same diffusion element, route1 and route 2 for Fe, with similar initial atomic environment as shown in TableS1 and the local diffusion environment as schematically shown in Fig.4(a1) and (b1). For route 1, when Fe diffuses to the saddle point, the up Mn atom change its spin direction and becomes FM. There are two competing factors that determine the change in Mn electronic structure, i.e., nearest-neighbor (NN) interaction and lattice volume [29]. In this case, the lattice distortion of Mn caused by the diffusion of Fe atoms is only 0.2 Å. From the LDOS of diffusion atom Fe and 1nn Mn at S2 site in Fig.4(b2) and (b3), we can see Mn adjusts its electronic structure to couple with Fe. In Fig.4(b1), before Fe diffusion, the two Mn atoms on the path are AFM state, in accordance with the Hund's rule. It is known that there is a strong magnetoelastic coupling in Fe-Mn alloys [34-36]. This magnetoelastic coupling is the result of the large spin-orbit coupling due to the large number of unpaired spins for both Fe and Mn and the close competing exchange between Mn-Mn and Mn-Fe. The coupling makes the local atoms reach a stable state, leading to a lower barrier. Note that, the FM-Mn also has a lower effective interaction volume than that of the AF-Mn [34], leaving a bigger "channel" for the diffusing Fe. This specificity phenomenon is mainly due to the flexibility of the Mn magnetic moment, which means an easy variation of the moment magnitude, to adapt different local environments. From TableS1, in the case of Co vacancy diffusions, one Mn atom around the diffusion path for route 3 rather than the others significantly reduce the barrier to 0.54 eV.

Therefore, our analysis on the process for the vacancy exchanging with its 1nn surrounding atoms shows the evidence that the presence of Mn is conducive to the diffusion of vacancies due to the "electron filling effect". Previous calculations on the diffusion of $3d$ solutes in Ni found a local minimum in the diffusion energy barrier profile at the position of the Mn solute with the occurrence of the maximum in the magnetic moment across the $3d$ row [37]. Recently, an in-situ TEM study on two quaternary Cantor alloys with different Mn-contents under heavy ion irradiation observed a big difference in dislocation loop growth kinetics, and speculated that the reason is due to Mn effect on the nucleation rate by increasing vacancy mobility [10]. In all, we show a proof-of-principle new mechanism that correlates element inducing point defect dynamic change with intrinsic macro-property behavior in this study. The addition of Mn with a unique half-filling $3d$-shell significantly accelerates the diffusion of vacancy by destroying the electron screen effect, resulting a large overlap of vacancy and interstitial diffusion, which contributes to the irradiation resistance.

This research was supported by the National Natural Science Foundation of China (12075044, 11705018, 51771015, 12075179) and as part of the Energy Dissipation to Defect Evolution (EDDE), an Energy Frontier Research Center funded by the US Department of Energy, Office of Science, Basic Energy Sciences under contract number DE-AC05-00OR22725.

*huangss@dlut.edu.cn, chenylu@xjtu.edu.cn

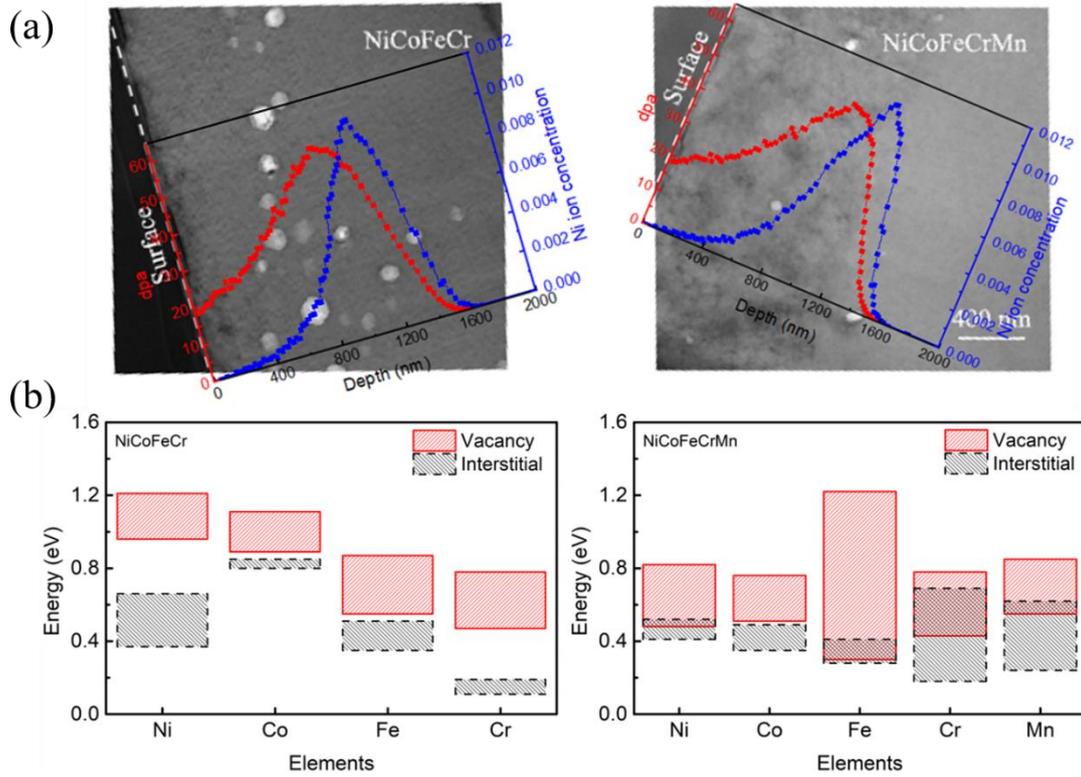

Fig.1 (a) The cross-section bright field TEM images of the alloys after 3 MeV $Ni^{2+}$ ion irradiation to a fluence of $5\times10^{16}$ ions/cm$^2$ at 853 K, the depth profiles of irradiation dose and implanted Ni ion concentration by SRIM code are also indicated. (b) The calculated migration barrier ranges of vacancy and interstitial. In each alloy, all possible diffusion paths for two random single-vacancy and more than fourteen interstitial atom diffusion paths are considered. Detailed diffusion pathways are summarized in Fig. S1-S4.

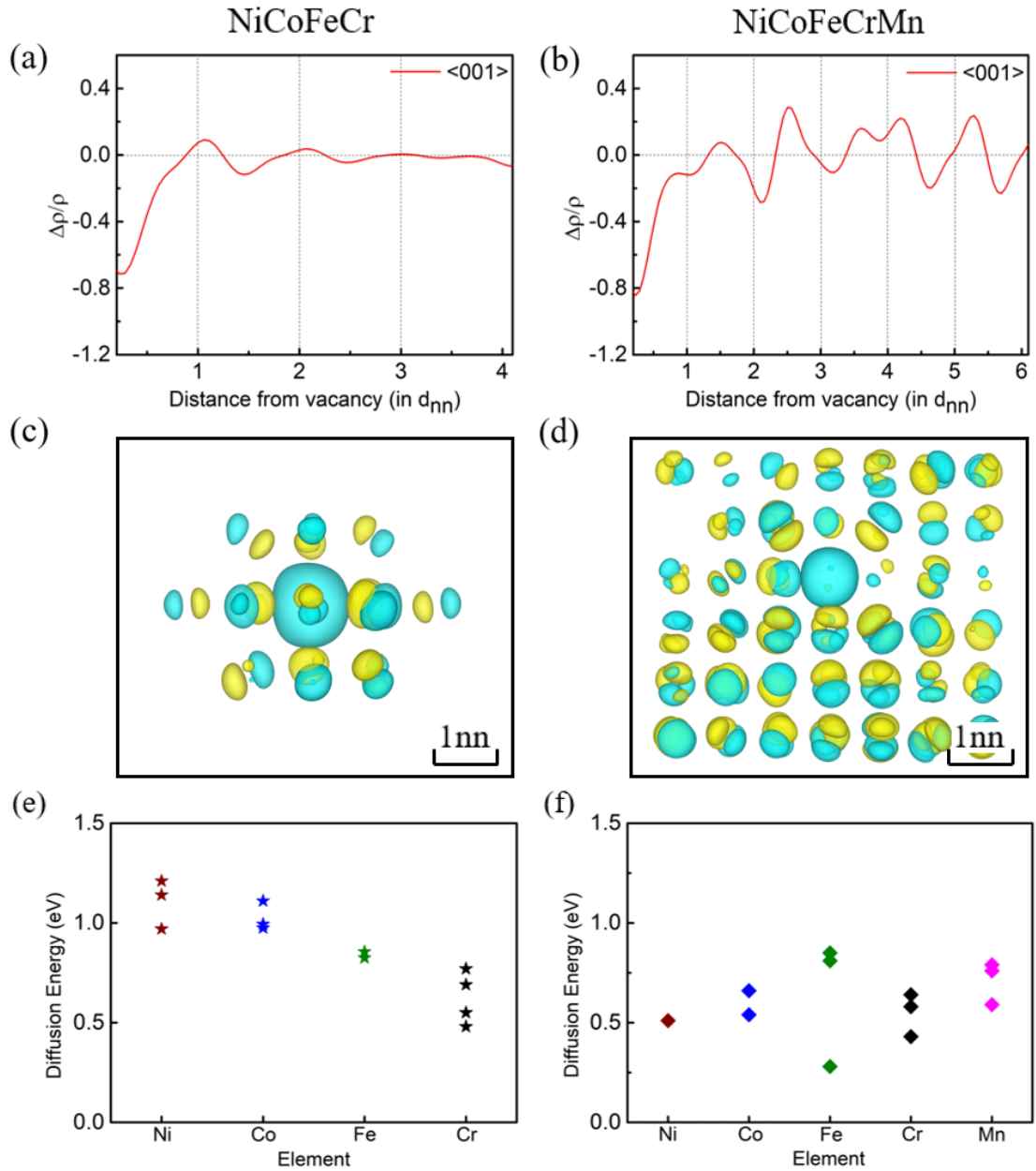

Fig.2 (a) and (b) are the charge density difference before and after vacancy formation along [001] directions in NiCoFeCr and NiCoFeCrMn alloy with (3x3x3) and (3x3x5) supercell respectively. Distance from the vacancy is in scale of the nearest-neighbor atom-atom distance $d_{nn}$=2.452 Å in NiCoFeCr and $d_{nn}$=2.471 Å in NiCoFeCrMn. (c) and (d) represent the calculated deformation charge density with the generation of a vacancy in the alloys with the same isosurface of 0.04 e/Å$^3$. (e) and (f) are the diffusion energy barrier of 1nn atoms exchange with the vacancy in the alloys.

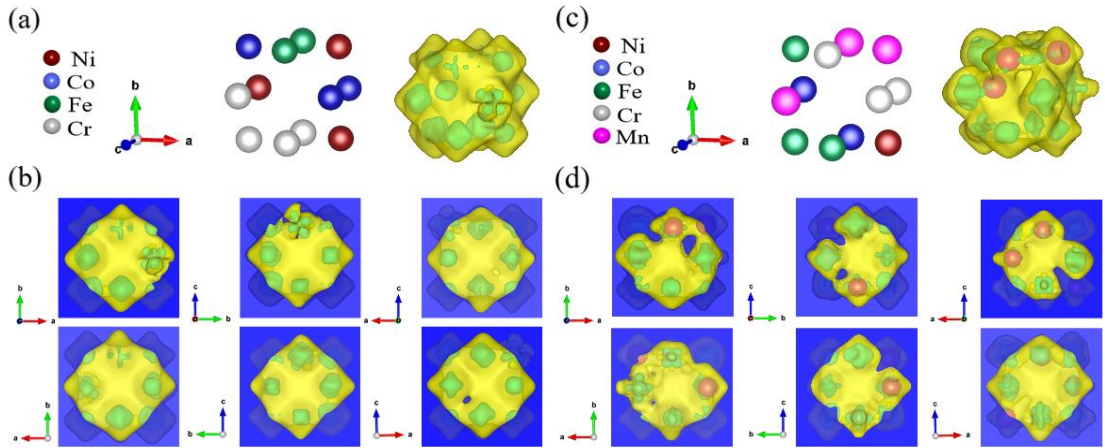

Fig. 3 (a) and (c) are the local atomic environment diagram and the deformation charge density of 1nn atoms around vacancy before and after vacancy formation in NiCoFeCr and NiCoFeCrMn alloy, respectively. (b) and (d) insert six lattice planes into the section of the deformation charge density in (a) and (c). The "broken" area with blue means "opening channel" with no electrons occupied through the two opposite plane. The isosurface is 0.016 e/Å$^3$. (For interpretation of the references to color in this figure legend, the reader is referred to the web version of this article.)

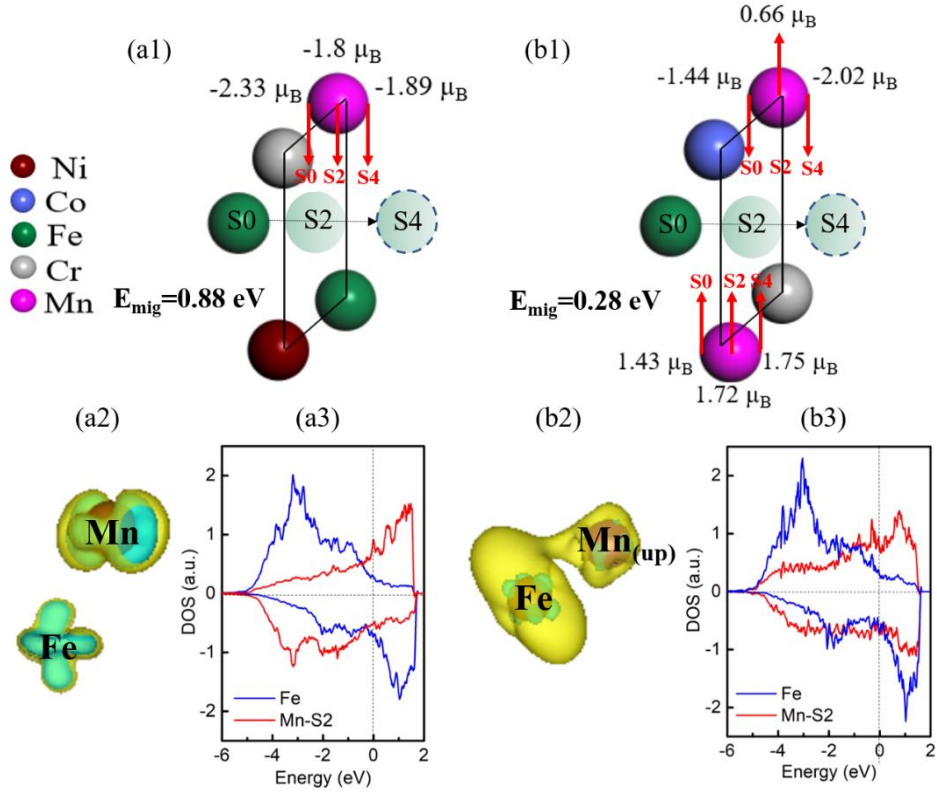

Fig.4 The diffusion process and corresponding Mn magnetic moments variation, deformation charge density and d-PDOS of Fe atom diffusion route 2 (a) and route 1 (b) in Table S1. (a1) and (b1) are the diffusion process and Mn magnetic moments variation from initial (S0) to saddle (S2) and final (S4) points. (a2), (b2) and (a3), (b3) present the deformation charge density and d-PDOS, respectively, of the diffusion atom Fe and 1nn Mn when Fe at S2. The DOS of Fe has no change during the diffusion process. Fermi level is located at 0 eV.

# Supplemental Material for "Tuned Vacancy Diffusion by Mn via Anomalous Friedel Oscillations in NiCoFeCrMn High Entropy Alloys"

## METHOD

First-principles density functional theory (DFT) calculations were performed based on the Projector Augmented Wave (PAW) method with the Perdew-Burke-Ernzerhof (PBE) exchange potential, as implemented in the Vienna Ab Initio Simulation Package (VASP) code [21]. The diffusion barriers and paths were investigated using the Climbing-Image Nudged Elastic Band (CI-NEB) method [22]. In this method, three intermediate images were used to optimize along the reaction path. The tetrahedron smearing method with Blöchl corrections [23] was used on fixed dimension/volume calculations to generate the density of states (DOS). Based on the ab initio ground-state wave function, we obtained five $3d$ orbital basis {$d_{xy}$, $d_{yz}$, $d_{z2}$, $d_{xz}$, $d_{x2-y2}$} for Mn atoms. The Brillouin zones were sampled using K point grids with a uniform spacing of $2\pi \times 0.04$ A$^{-1}$. The model structures were fully optimized by using thresholds of $10^{-4}$ eV and 0.02 eV/Å for the total energy and force, respectively. The electron wavefunctions for quaternary and quinary alloys were expanded by the plane wave basis up to 400 eV, while those for ternary alloys were expanded to 300 eV. The preceding parameters were all carefully selected through pre-calculation to ensure that the results were accurate and that computational resource use was minimized as much as possible.

The structures were modeled utilizing the similar local approximation environment (SAE) [17,18], which were generated by creating similar local atomic environments for all lattice sites. In this study, the CoCrNiFe quaternary alloy and the CoCrNiFeMn quinary HEA are 3×3×3 supercells with a total equivalent atomic number of 108 and 3×3×5 supercell with a total equivalent atomic number of 180, respectively.

## DIFFUSION PATH

Figure S1 and S2 are the NEB calculations of vacancy migration barriers based on different element types in a 107-atom NiCoFeCr and a 179-atom NiCoFeCrMn supercell. We choose Co vacancy in NiCoFeCr and NiCoFeCrMn alloy as initial position (S0), and then make 1nn atoms diffuse to the vacancy. Figure S3 and S4 are the NEB calculation of self-interstitial migration barriers based on different element types in a 109-atom NiCoFeCr and 181-atom NiCoFeCrMn supercell. Interstitials diffuse from initial site (S0) to final site (S4). S0 and S4 are two adjacent stable interstitial positions. The energy difference between the saddle point position (the position with the highest energy) and the initial position is the diffusion barrier. We summarize all the diffusion energies and the atomic environments of diffusion atoms at S0 position in Table S1.

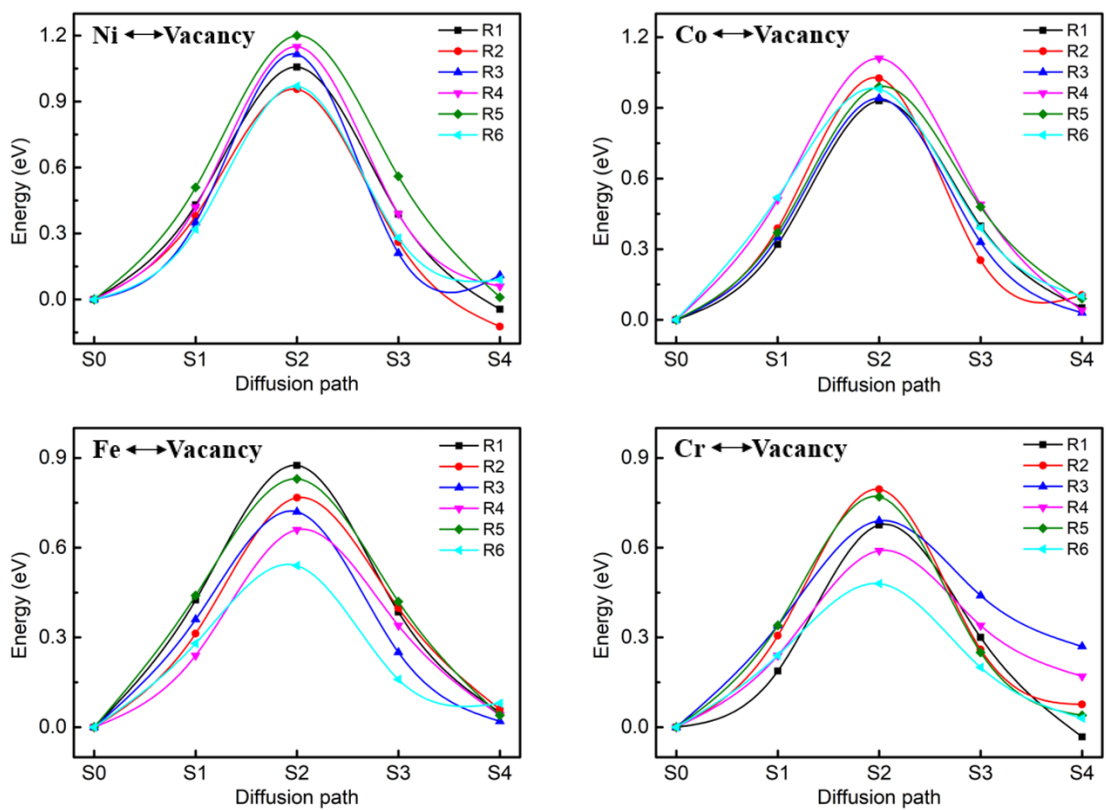

Fig. S1 NEB calculations of vacancy migration barriers based on different element types in a 107-atom NiCoFeCr supercell. Vacancies diffuse from the initial site (S0) to the final site (S4).

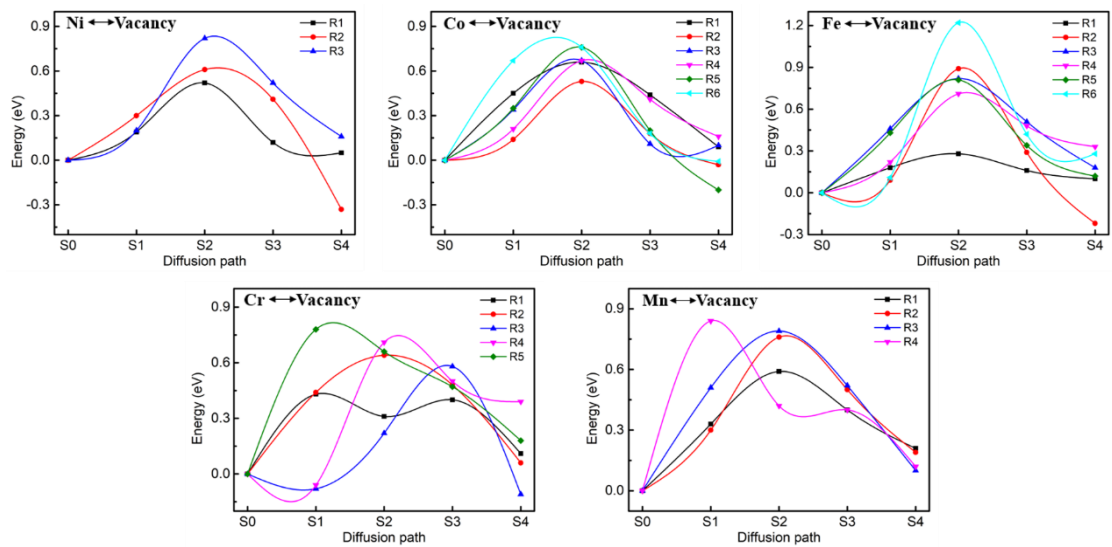

Fig. S2 NEB calculations of vacancy migration barriers based on different element types in a 179-atom NiCoFeCrMn supercell. Vacancies diffuse from the initial site (S0) to the final site (S4).

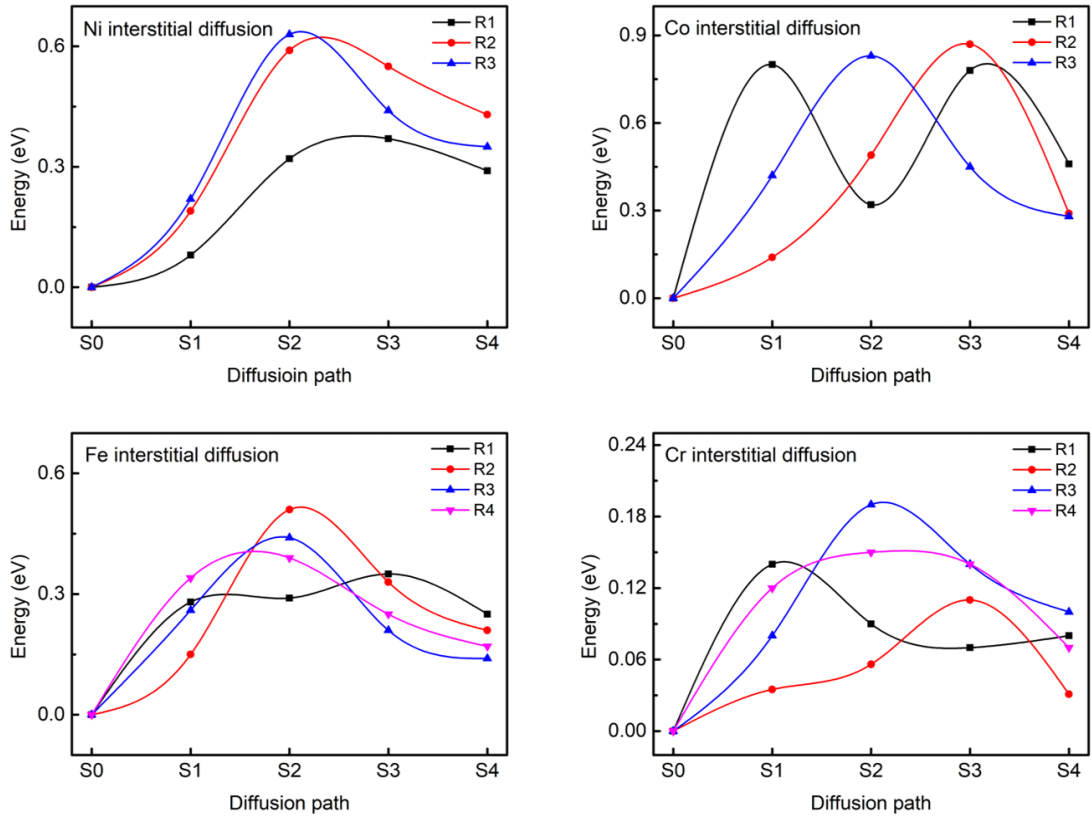

Fig. S3 NEB calculation of self-interstitial migration barriers based on different element types in a 109-atom NiCoFeCr supercell. Interstitials diffuse from initial site (S0) to final site (S4).

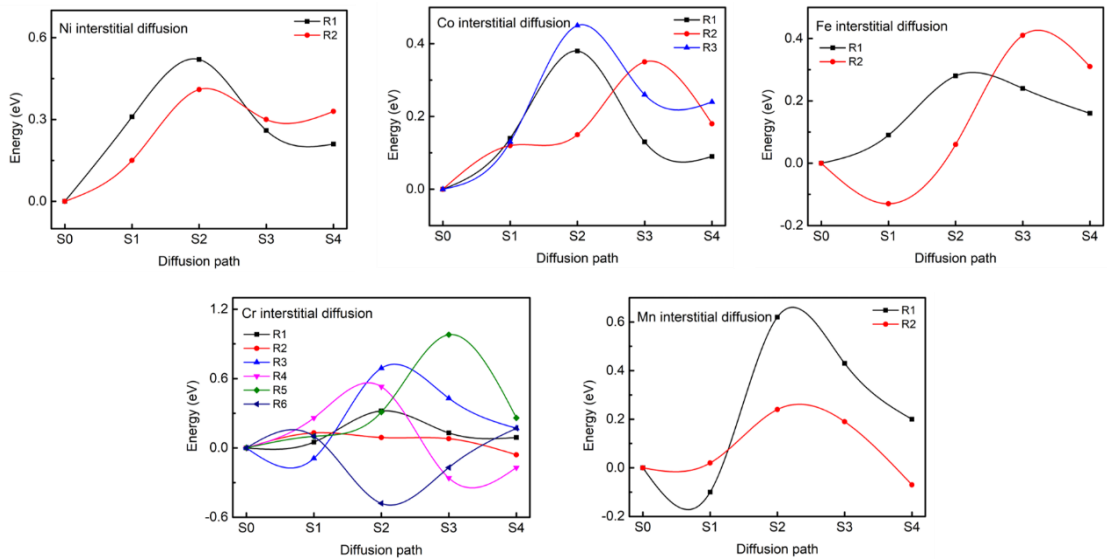

Fig. S4 NEB calculation of self-interstitial migration barriers based on different element types in a 181-atom NiCoFeCrMn supercell. Interstitials diffuse from initial site (S0) to final site (S4).

Table S1 The diffusion energy barrier for the elements exchanging with a vacancy in NiCoFeCrMn. Routes with different local 1nn atomic environment for the diffusing elements are presented. Symbol * means the number of Mn atoms in the diffusion path.

| Diffusing elements | Vacancy | Diffusion route | 1nn atomic environment for the diffusing element | | | | | Diffusion energy barrier (eV) |
|---|---|---|---|---|---|---|---|---|
| | | | Fe | Ni | Co | Cr | Mn | |
| Fe | V1 | R1 | 2 | 0 | 2 | 3 | 4(2*) | 0.28 |
| | | R2 | 2 | 2 | 2 | 3 | 2(1*) | 0.88 |
| | | R3 | 3 | 2 | 3 | 1 | 2(1*) | 0.82 |
| | V2 | R4 | 3 | 2 | 4 | 0 | 2(0*) | 0.71 |
| | | R5 | 1 | 5 | 3 | 1 | 1(0*) | 0.81 |
| | | R6 | 1 | 3 | 5 | 2 | 0(0*) | 1.22 |
| Co | V1 | R1 | 6 | 1 | 3 | 0 | 1(1*) | 0.54 |
| | | R2 | 3 | 4 | 2 | 1 | 1(0*) | 0.66 |
| | V2 | R3 | 2 | 1 | 4 | 3 | 1(0*) | 0.67 |
| | | R4 | 3 | 3 | 3 | 1 | 1(0*) | 0.67 |
| | | R5 | 3 | 2 | 2 | 3 | 1(0*) | 0.76 |
| | | R6 | 5 | 2 | 2 | 1 | 1(0*) | 0.76 |
| Cr | V1 | R1 | 1 | 1 | 1 | 4 | 4(2*) | 0.43 |
| | | R2 | 2 | 2 | 0 | 4 | 3(1*) | 0.64 |
| | | R3 | 1 | 4 | 3 | 1 | 2(2*) | 0.58 |
| | V2 | R4 | 1 | 1 | 3 | 4 | 2(1*) | 0.71 |
| | | R5 | 2 | 0 | 3 | 4 | 2(1*) | 0.78 |
| Ni | V1 | R1 | 3 | 4 | 1 | 2 | 1(0*) | 0.51 |
| | V2 | R2 | 3 | 2 | 2 | 0 | 4(2*) | 0.61 |
| | | R3 | 2 | 3 | 2 | 2 | 2(1*) | 0.82 |
| Mn | V1 | R1 | 2 | 1 | 4 | 2 | 2(1*) | 0.59 |
| | | R2 | 4 | 1 | 2 | 4 | 0(0*) | 0.76 |
| | | R3 | 1 | 1 | 1 | 4 | 4(1*) | 0.79 |
| | V2 | R4 | 4 | 1 | 1 | 4 | 1(0*) | 0.84 |

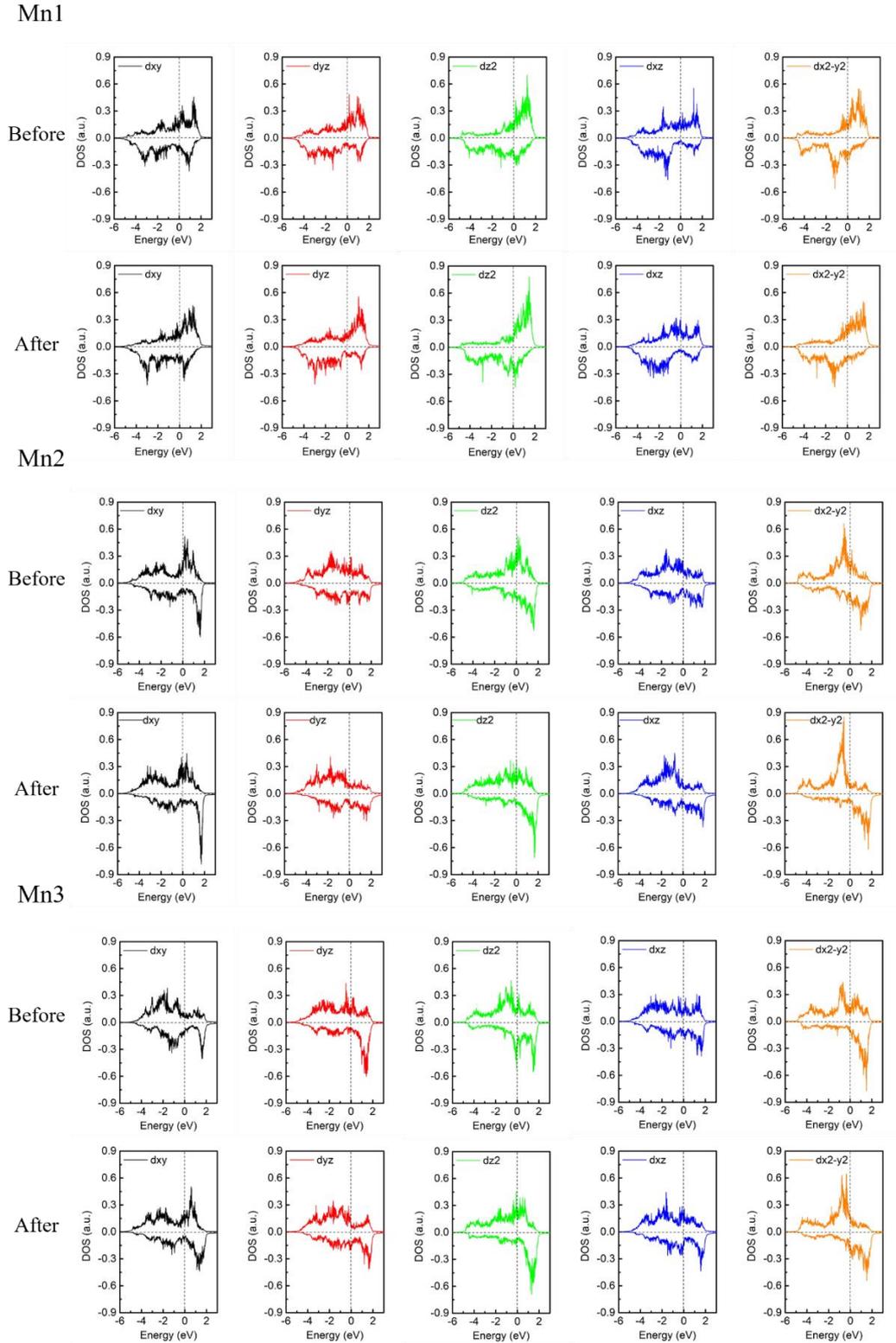

Fig. S5 d-orbital splitting of various states (dxy, dyz, $dz^2$, dxz, and $dx^2$-$y^2$) for Mn atoms (Mn1, Mn2, and Mn3) around the Co vacancy before and after the introduction of vacancy in NiCoFeCrMn alloy. Fermi level is located at 0 eV.